\documentclass[aps,preprintnumbers,nofootinbibt,referee, keywords]{revtex4}
\usepackage{graphicx}
\usepackage{amsmath}
\usepackage{bm}
\usepackage{color}

\def\be{\begin{equation}}
\def\ee{\end{equation}}
\def\bea{\begin{eqnarray}}
\def\eea{\end{eqnarray}}

\begin{document}

\title{Exact analytical solutions of the Susceptible-Infected-Recovered (SIR)
epidemic model and  of the SIR model with equal death and
birth rates}
\author{Tiberiu Harko}
\email{t.harko@ucl.ac.uk}
\affiliation{Department of Mathematics, University College London, Gower Street, London
WC1E 6BT, United Kingdom}
\author{Francisco S. N. Lobo}
\email{flobo@cii.fc.ul.pt}
\affiliation{Centro de Astronomia e Astrof\'{\i}sica da Universidade de Lisboa, Campo
Grande, Edific\'{\i}o C8, 1749-016 Lisboa, Portugal}
\author{M. K. Mak}
\email{mkmak@vtc.edu.hk}
\affiliation{Department of Computing and Information Management, Hong Kong Institute of
Vocational Education, Chai Wan, Hong Kong, P. R. China}

\begin{abstract}
In this paper, the exact analytical solution of the
Susceptible-Infected-Recovered (SIR) epidemic model is obtained in a
parametric form. By using the exact solution we investigate some explicit
models corresponding to fixed values of the parameters, and show that the
numerical solution reproduces exactly the analytical solution. We also show
that the generalization of the SIR model, including births and deaths,
described by a nonlinear system of differential equations, can be reduced to
an Abel type equation. The reduction of the complex SIR model with vital
dynamics to an Abel type equation can greatly simplify the analysis of its
properties. The general solution of the Abel equation is obtained by using a
perturbative approach, in a power series form, and it is shown that the
general solution of the SIR model with vital dynamics can be represented in
an exact parametric form.

\end{abstract}

\maketitle


\section{Introduction}

The outbreak and spread of diseases has been questioned and studied for many
years. In fact, John Graunt was the first scientist who attempted to
quantify causes of death systematically \cite{1}, and his analysis of causes
of death ended up with a theory that is now well established among modern
epidemiologists. Daniel Bernoulli was the first mathematician to propose a
mathematical model describing an infectious disease. In 1760 he modelled the
spread of smallpox \cite{Ber1}, which was prevalent at the time, and argued
the advantages of variolation \cite{Ber2}. A simple deterministic
(compartmental) model predicting the behavior of epidemic outbreaks was
formulated by A. G. McKendrick and W. O Kermack in 1927 \cite{2}. In their
mathematical epidemic model, called the Susceptible-Infected-Recovered (SIR)
model, or the $xyz$ model, to describe the spread of diseases, McKendrick
and Kermack proposed the following nonlinear system of ordinary differential
equations \cite{2}
\begin{equation}
\frac{dx}{dt}=-\beta x\left( t\right) y\left( t\right) ,  \label{A1}
\end{equation}%
\begin{equation}
\frac{dy}{dt}=\beta x\left( t\right) y\left( t\right) -\gamma y\left(
t\right) ,  \label{A2}
\end{equation}%
and
\begin{equation}
\frac{dz}{dt}=\gamma y\left( t\right) ,  \label{A3}
\end{equation}%
respectively, with the initial conditions $x\left( 0\right) =N_{1}\geq 0$, $%
y\left( 0\right) =N_{2}\geq 0$ and $z\left( 0\right) =N_{3}\geq 0$, $%
N_{i}\in \Re $, $i=1,2,3$, and where the infection rate $\beta $ and the
mean recovery rate $\gamma $ are positive constants. In this model a fixed
population with only three compartments is considered: susceptible (S) $%
x\left( t\right) $, infected (I) $y\left( t\right) $, and recovered (R) $%
z\left( t\right) $, respectively. The compartments used for this model
consist of three classes:

a) $x\left( t\right) $ represents the number of individuals not yet infected
with disease at time $t$ or those susceptible to the disease,

b) $y\left( t\right) $ denotes the number of individuals who have been
infected with the disease, and are capable of spreading the disease to those
in the susceptible category, and

c) $z\left( t\right) $ is the compartment of the individuals who have been
infected and then recovered from the disease. Those in this category are not
able to be infected again, or to transmit the infection to others. For this
model the initial conditions $x(0)=N_1$, $y(0)=N_2$ and $z(0)=N_3$ are not
independent, since they must satisfy the condition $N_1+N_2+N_3=N$, where $N$
is the total fixed number of the individuals in the given population. The
constants $\beta $ and $\gamma $ give the transition rates between
compartments. The transition rate between S (Susceptible) and I (infected)
is $\beta $ I, where $\beta $ is the contact rate, which takes into account
the probability of getting the disease in a contact between a susceptible
and an infectious subject \cite{14,Mur, 15, Mur1}. The transition rate
between I (Infected) and R (recovered), is $\gamma $, which has the meaning
of the rate of recovery or death. If the duration of the infection is
denoted $D$, then $\gamma = 1/D$, since an individual experiences one
recovery in $D$ units of time \cite{14,Mur, 15, Mur1}. Since $\beta $ and $%
\gamma $ are interpreted as transition rates (probabilities), their range is
$0\leq \beta \leq 1$ and $0\leq \gamma \leq 1$, respectively.

In many infectious diseases, such as in the case of measles, there is an
arrival of new susceptible individuals into the population. For this type of
situation deaths must be included in the model. By considering a population
characterized by a death rate $\mu $ and a birth rate equal to the death
rate, the non-linear system of the differential equations representing this
epidemic model is given by \cite{14,Mur,15,Mur1},
\begin{equation}
\frac{dx}{dt}=-\beta xy+\mu \left( N-x\right) ,  \label{K1}
\end{equation}%
\begin{equation}
\frac{dy}{dt}=\beta xy-\left(\gamma +\mu \right)y,  \label{K2}
\end{equation}%
\begin{equation}
\frac{dz}{dt}=\gamma y-\mu z.  \label{K3}
\end{equation}%
and it must be considered with the initial conditions $x(0)=N_1$, $y(0)=N_2$
and $z(0)=N_3$, with the constants $N_i$, $i=1,2,3$ satisfying the condition
$\sum _{i=1}^3{N_i}=N$.

The non-linear differential equations systems, given by Eqs.~(\ref{A1})-(\ref%
{A3}) and (\ref{K1})-(\ref{K3}) represent modified three-dimensional
competitive Lotka--Volterra type models \cite{14,15}. These systems can also
be related to the so-called T systems, introduced recently in \cite{Tigan},
and which have the form
\begin{equation}
\frac{dx}{dt}=a_0(y-x),
\end{equation}
\begin{equation}
\frac{dy}{dt}=(c_0-a_0)x-a_0xz,
\end{equation}
\begin{equation}
\frac{dz}{dt}=-b_0z+xy,
\end{equation}
which is chaotic for $a_0=2.1$, $b_0=0.6$ and $c_0=30$. The mathematical
properties of the T-system were studied in \cite{Tigan1,Tigan2,Tigan3}.

In recent years, the mathematical epidemic models given by Eqs.~(\ref{A1})-(%
\ref{A3}) and (\ref{K1})-(\ref{K3}) were investigated numerically in a number of papers, with the use of a wide
variety of methods and techniques, namely, the Adomian decomposition method %
\cite{5}, the variational iteration method \cite{6}, the homotopy
perturbation method \cite{7}, and the differential transformation method %
\cite{8}, respectively. Very recently, a stochastic epidemic-type model with
enhanced connectivity was analyzed in \cite{3}, and an exact solution of the
model was obtained. With the use of a quantum mechanical approach the master
equation was transformed via a quantum spin operator formulation. The
time-dependent density of infected, recovered and susceptible populations
for random initial conditions was calculated exactly. An exact solution of a
particular case of the SIR and Susceptible--Infected--Susceptible (SIS)
epidemic models was presented in \cite{4}. Note that in the latter SIS
model, the infected return to the susceptible class on recovery because the
disease confers no immunity against reinfection. A SIR epidemic model with
nonlinear incidence rate and time delay was investigated in \cite{n3}, while
an age-structured SIR epidemic model with time periodic coefficients was
studied in \cite{n2}. In fact, the standard pair approximation equations for
the Susceptible-Infective-Recovered-Susceptible (SIRS) model of infection
spread on a network of homogeneous degree $k$ predict a thin phase of
sustained oscillations for parameter values that correspond to diseases that
confer long lasting immunity. Indeed, the latter SIRS model has been
thoroughly studied, with the results strongly suggesting that its
stochastic Markovian version does not yield sustained oscillations \cite{Nunes1}. A stochastic model of infection dynamics based on the Susceptible-Infective-Recovered  model, where the distribution of the recovery times can be tuned, interpolating between exponentially distributed recovery times, as in the standard SIR model, and recovery after a fixed infectious period, was investigated in \cite{Nunes2}. For large populations, the spectrum of fluctuations around the deterministic limit of the model was obtained analytically.

An epidemic model with stage structure was introduced in \cite{n5}, with the
period of infection partitioned into the early and later stages according to
the developing process of infection. The basic reproduction number of this
model is determined by the method of next generation matrix. The global
stability of the disease-free equilibrium and the local stability of the
endemic equilibrium have been obtained, with the global stability of the
endemic equilibrium is determined under the condition that the infection is
not fatal. Lyapunov functions for classical SIR and SIS epidemiological
models were introduced in \cite{n1}, and the global stability of the endemic
equilibrium states of the models were thereby established. A new Lyapunov
function for a variety of SIR and SIRS models in epidemiology was introduced
in \cite{n4}. Traveling wave trains in generalized two-species predator-prey
models and two-component reaction-diffusion equations were considered in %
\cite{Mancas3}, and the stability of the fixed points of the traveling wave
ordinary differential equations in the usual ''spatial'' variable was
analyzed. For general functional forms of the nonlinear prey birthrate/prey
death rate or reaction terms, a Hopf bifurcation occurs at two different
critical values of the traveling wave speed. Subcritical Hopf bifurcations
yield more complex post-bifurcation dynamics in the wavetrains of the system
of the partial differential equations. In order to investigate the
post-bifurcation dynamics all the models were integrated numerically, and
chaotic regimes were characterized by computing power spectra,
autocorrelation functions, and fractal dimensions, respectively.

It is the purpose of this paper to present the exact analytical solution of
the SIR epidemic model. The solution is obtained in an exact parametric
form. The generalization of the SIR model including births deaths, described
by Eqs.~(\ref{K1})--(\ref{K3}), is also considered, and we show that the
nonlinear system of differential equations governing the generalized SIR
model can be reduced to an Abel type equation. The general solution of the
Abel equations is obtained by using an iterative method and, once the
solution of this ordinary differential equation is known, the general
solution of the SIR model with vital dynamics can be obtained, similarly to
the standard SIR model, in an exact parametric form.

The present paper is organized as follows. The exact solution of the SIR
epidemic model is presented in Section \ref{II}. The nonlinear system of
differential equations governing the SIR model with deaths is reduced to an
Abel type equation, and the general solution of the model equations is
obtained in an exact parametric form in Section \ref{III}. We conclude our
results in Section \ref{IV}.

\section{The exact solution of the SIR epidemic model}

\label{II}

By adding Eqs.~(\ref{A1})--(\ref{A3}), yields the following differential
equation,
\begin{equation}
\frac{d}{dt}\left[ x(t)+y(t)+z(t)\right] =0,  \label{A4}
\end{equation}%
which can be immediately integrated to give
\begin{equation}
x(t)+y(t)+z(t)=N,\forall t\geq 0,  \label{A5}
\end{equation}%
where $x(t)>0$, $y(t)>0$ and $z(t)>0$, $\forall t\geq 0$. Hence, the total
population $N=N_{1}+N_{2}+N_{3}$ must be an arbitrary positive integration
constant. This is consistent with the model in which only a fixed population
$N$ with only three compartments is considered.

\subsection{The general evolution equation for the SIR model}

As a next step in our study, we differentiate Eq.~(\ref{A1}) with respect to
the time $t$, thus obtaining the following second order differential
equation,
\begin{equation}
\frac{dy}{dt}=-\frac{1}{\beta }\left[ \frac{x^{\prime \prime }}{x}-\left(
\frac{x^{\prime }}{x}\right) ^{2}\right] ,  \label{A6}
\end{equation}
where the prime represents the derivative with respect to time $t$. By
inserting Eqs.~(\ref{A1}) and (\ref{A6}) into Eq.~(\ref{A2}), the latter is
transformed into
\begin{equation}
\frac{x^{\prime \prime }}{x}-\left( \frac{x^{\prime }}{x}\right) ^{2}+\gamma
\frac{x^{\prime }}{x}-\beta x^{\prime }=0.  \label{A7}
\end{equation}

By eliminating $y$ from Eqs. (\ref{A1}) and (\ref{A3}) yields
\begin{equation}
\frac{dz}{dt}=-\frac{\gamma }{\beta }\left( \frac{x^{\prime }}{x}\right) \,.
\label{A8}
\end{equation}
which can be integrated to give
\begin{equation}
x=x_{0}e^{-\frac{\beta }{\gamma }z},  \label{A82}
\end{equation}
where $x_{0}$ is a positive integration constant. By estimating Eq.~(\ref%
{A82}) at $t=0$ provides the following value for the integration constant
\begin{equation}
x_0=N_1e^{\frac{\beta }{\gamma }N_3}.
\end{equation}
From Eq.~(\ref{A82}), it is easy to obtain the relation
\begin{equation}
x^{\prime }=-\frac{x_{0}\beta }{\gamma }z^{\prime }e^{-\frac{\beta }{\gamma }%
z}.  \label{A83}
\end{equation}

Now, differentiation of Eq.~(\ref{A8}) with respect to the time $t$ leads to
the second order differential equation
\begin{equation}
z^{\prime \prime }=-\frac{\gamma }{\beta }\left[ \frac{x^{\prime \prime }}{x}%
-\left( \frac{x^{\prime }}{x}\right) ^{2}\right] .  \label{A9}
\end{equation}

By inserting Eqs.~(\ref{A8}), (\ref{A83}) and (\ref{A9}) into Eq.~(\ref{A7}%
), the latter becomes the basic differential equation describing the spread
of a non-fatal disease in a given population
\begin{equation}
z^{\prime \prime }=x_{0}\beta z^{\prime }e^{-\frac{\beta }{\gamma }z}-\gamma
z^{\prime }.  \label{A10}
\end{equation}

Eq.~(\ref{A10}) is equivalent to the system of differential equations Eqs.~(%
\ref{A1})--(\ref{A3}), respectively.

\subsection{The general solution of the evolution equation of the SIR model}

In order to solve the nonlinear differential equation Eq.~(\ref{A10}), we
introduce a new function $u\left( t\right) $ defined as
\begin{equation}
u=e^{-\frac{\beta }{\gamma }z}.  \label{u}
\end{equation}
At $t=0$, $u$ has the initial value
\begin{equation}
u(0)=u_0=e^{-\frac{\beta }{\gamma }N_3}.
\end{equation}

Substituting Eq.~(\ref{u}) into Eq.~(\ref{A10}), we obtain the following
second order differential equation for $u$,
\begin{equation}
u\frac{d^{2}u}{dt^{2}}-\left( \frac{du}{dt}\right) ^{2}+\left( \gamma
-x_{0}\beta u\right) u\frac{du}{dt}=0.  \label{A11}
\end{equation}
Next we introduce the new function $\phi $, defined as
\begin{equation}
\phi =\frac{dt}{du}.  \label{ph}
\end{equation}

With the help of the transformation given by Eq.~(\ref{ph}), Eq.~(\ref{A11})
becomes a Bernoulli type differential equation,
\begin{equation}
\frac{d\phi }{du}+\frac{1}{u}\phi =\left( \gamma -x_{0}\beta u\right) \phi
^{2},  \label{B}
\end{equation}%
with the general solution given by
\begin{equation}
\phi =\frac{1}{u\left( C_{1}-\gamma \ln u+x_{0}\beta u\right) },  \label{B1}
\end{equation}%
where $C_{1}$ is an arbitrary integration constant. In view of Eqs.~(\ref{ph}%
) and (\ref{B1}), we obtain the integral representation of the time as
\begin{equation}
t-t_{0}=\int_{u_{0}}^{u}\frac{d\xi }{\xi \left( C_{1}-\gamma \ln \xi
+x_{0}\beta \xi \right) },  \label{t}
\end{equation}%
where $t_{0}$ is an arbitrary integration constant, and one may choose $%
t_{0}=0$, without loss of generality. Hence we have obtained the complete
exact solution of the system of Eqs.~(\ref{A1})--(\ref{A3}), describing the
SIR epidemic model, given in a parametric form by
\begin{equation}
x=x_{0}u,  \label{s1}
\end{equation}%
\begin{equation}
y=\frac{\gamma }{\beta }\ln u-x_{0}u-\frac{C_{1}}{\beta },  \label{s2}
\end{equation}%
\begin{equation}
z=-\frac{\gamma }{\beta }\ln u,  \label{s3}
\end{equation}%
with $u$ taken as a parameter. Now adding Eqs.~(\ref{s1}), (\ref{s2}) and (%
\ref{s3}), we obtain
\begin{equation}
x+y+z=-\frac{C_{1}}{\beta }.  \label{xyz}
\end{equation}

Comparing Eq.~(\ref{xyz}) with Eq.~(\ref{A5}), we have
\begin{equation}  \label{C1}
C_1=-\beta N,
\end{equation}
and hence $C_{1}$ is a negative integration constant. Eqs.~(\ref{t})-(\ref%
{C1}) give the exact parametric solution of the SIR system of three
differential equations, with $u$ taken as parameter. The solution describes
exactly the dynamical evolution of the SIR system for any given initial
conditions $x(0)=N_1$, $y(0)=N_2$ and $z(0)=N_3$, and for arbitrary values
of $\beta $ and $\gamma $. The numerical values of the two constants in the
solution, $u_0$ and $C_1$ are determined by the model parameters and the
initial conditions. Any change in the numerical values of the initial
conditions and/or of the rate parameters will not affect the validity of the
solution.

In order to compare the results of the present analytical solution with the
results of the numerical integration of the system of differential equations
Eqs.~(\ref{A1})-(\ref{A3}) we adopt the initial values and the numerical
values for the coefficients considered in \cite{8}. Hence we take $N_{1}=20$%
, $N_{2}=15$, and $N_{3}=10$, respectively. For the parameter $\beta $ we
take the value $\beta =0.01$, while $\gamma =0.02$. The variations of $x(t)$%
, $y(t)$ and $z(t)$ obtained by both numerical integration and the use of
the analytical solution are represented, as a function of time, in Fig.~\ref%
{fig1}.

\begin{figure}[tbp]
\centering \includegraphics[width=100mm, height=70mm]{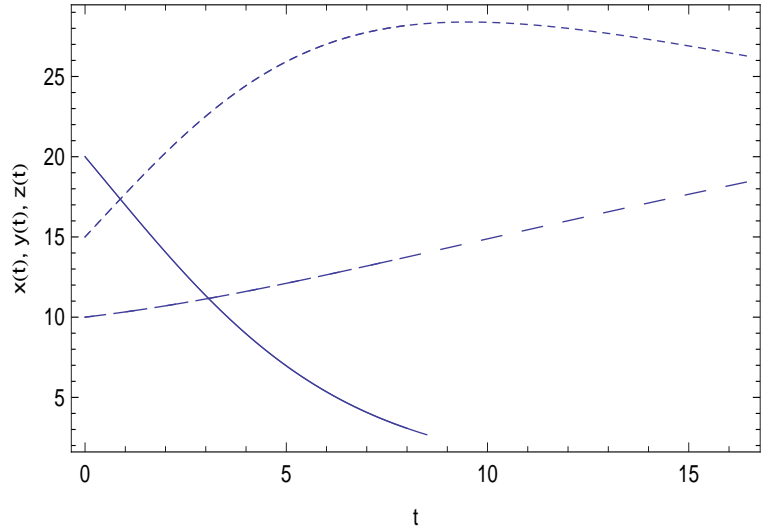}
\caption{ Variation of $x(t)$ (solid curve), $y(t)$ (dotted curve) and $z(t)$
(dashed curve), obtained by the numerical integration of the differential
equations Eqs.~(\ref{A1})--(\ref{A3}), and with the use of the analytical
solution, for $\protect\beta =0.01$ and $\protect\gamma =0.02$. The initial
conditions are $x(0)=N_1=20$, $y(0)=N_2=15 $, and $z(0)=N_3=10$,
respectively. The numerical and the analytical solutions completely overlap.}
\label{fig1}
\end{figure}

\begin{figure}[tbp]
\centering \includegraphics[width=100mm, height=70mm]{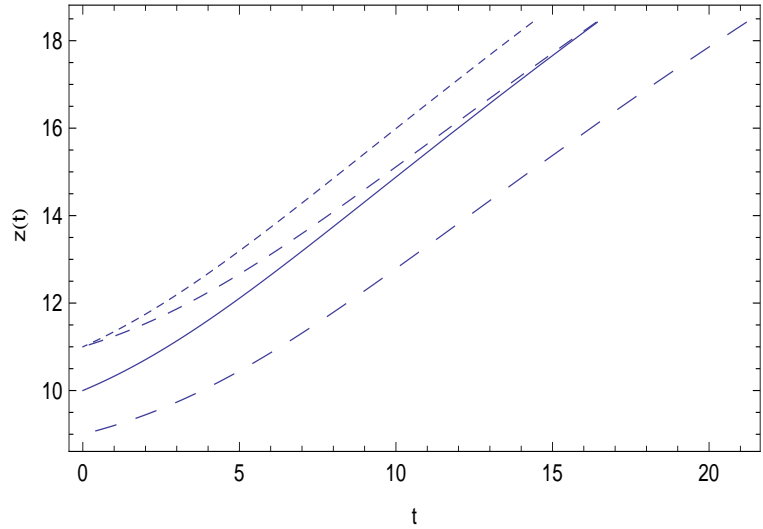}
\caption{ Time variation of $z(t)$ obtained with the use of the analytical
solution, for $\protect\beta =0.01$ and $\protect\gamma =0.02$, and for
different initial conditions: $x(0)=N_1=20$, $y(0)=N_2=15 $, and $%
z(0)=N_3=10 $ (solid curve), $x(0)=N_1=19$, $y(0)=N_2=16 $, and $z(0)=N_3=11$
(dotted curve), $x(0)=N_1=22$, $y(0)=N_2=11 $, and $z(0)=N_3=11$ (dashed
curve), and $x(0)=N_1=24$, $y(0)=N_2=9$, and $z(0)=N_3=9$ (long dashed
curve), respectively. }
\label{fig2}
\end{figure}

As one can see from the figure, the analytical solution perfectly reproduces
the results of the numerical integration. The exact solution we found here
is also in complete agreement with the numerical results obtained in \cite{5}%
-\cite{8}. The variation of $z(t)$ for different initial conditions $x(0)$, $%
y(0)$ and $z(0)$ is represented in Fig.~\ref{fig2}.

In the present Section, after a brief discussion of the general properties
of the SIR model with births and deaths, we will show that the time
evolution of the SIR model with equal birth and death rates can be obtained
from the study of a single first order Abel type differential equation. An
iterative approach for solving this equation is also presented.

\section{The SIR model with equal death and birth rates}

\label{III}

In the present Section we consider the extension of the simple SIR model
given by Eqs.~(\ref{A1})-(\ref{A3}) by including equal rates of births and
deaths. In this case the system of differential equations we are going to
consider is given by Eqs.~(\ref{K1})-(\ref{K3}).

It is easy to see that by adding Eqs.~(\ref{K1})--(\ref{K3}), and
integrating the resulting equation yields the following result
\begin{equation}
x(t)+y(t)+z(t)=N+N_{0}e^{-\mu t},
\end{equation}%
for this model, where $N_{0}$ is an arbitrary integration constant. In order
that to total number of individuals is a constant,
\begin{equation}
x(t)+y(t)+z(t)=N,\forall t\geq 0,
\end{equation}
we must fix the arbitrary integration constant $N_0$ as zero, $N_{0}=0$.

\begin{figure}[tbp]
\centering \includegraphics[width=100mm, height=70mm]{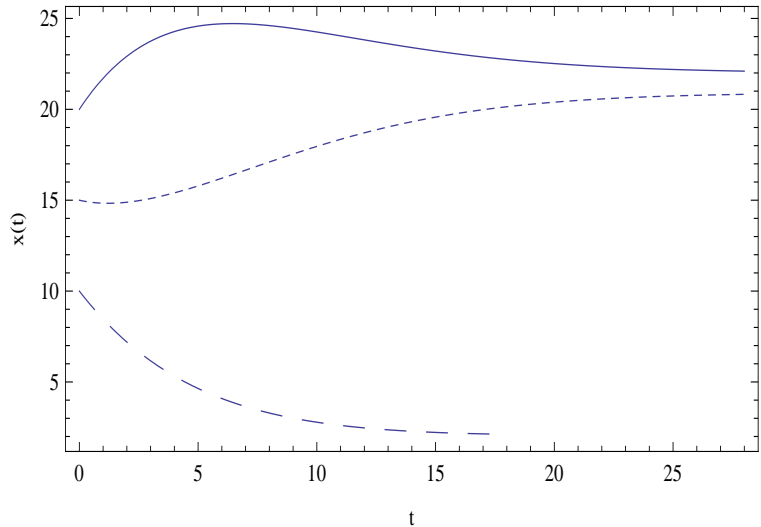}
\caption{ Variation of $x(t)$ (solid curve), $y(t)$ (dotted curve) and $z(t)$
(dashed curve), obtained by the numerical integration of the differential
equations Eqs.~(\ref{K1})-(\ref{K3}) of the SIR model with equal death and
birth rates for $\protect\beta =0.01$, $\protect\gamma =0.02$, and $\protect%
\mu =0.20$. The initial conditions are $x(0)=N_1=20$, $y(0)=N_2=15 $, and $%
z(0)=N_3=10$, respectively. }
\label{fig3}
\end{figure}

The variation of $x(t)$, $y(t)$ and $z(t)$ for the SIR model with vital
dynamics is represented in Fig.~\ref{fig3}.

\subsection{Qualitative properties of the SIR model with vital dynamics}

Both the simple SIR model ($\mu=0$) and the SIR model with vital dynamics ($%
\mu \neq 0$) are two dimensional dynamical systems in the $x+y+z=N$
invariant plane. There is no chaotic behaviour in the plane -- essentially
because the existence and unicity theorem prevents (in dimension 2) the
existence of transversal homoclinic points. The dynamics of both systems is
simple, and well understood, including the bifurcation that takes place at $%
\beta N=\gamma + \mu $. There is also no chaotic behaviour for $\mu \neq 0$
outside the $x+y+z=N$ invariant plane, because, as one can see from Eqs.~(%
\ref{K1})--(\ref{K3}), for the time evolution of $x+y+z$, trajectories with
initial conditions out of the invariant plane tend exponentially fast to the
invariant plane -- there can be no attractor or invariant set, chaotic or
otherwise, outside the invariant plane. For $\mu=0$, $y=0$ is a line of
degenerate equilibria, for all parameter values.

For $\mu \neq 0$ and $\beta <0$, $x=N,y=0,z=0$ is a global attractor -- a
stable node --, while for $\beta >0$,
\begin{equation}
x^{*}=\frac{\gamma + \mu}{\beta},
\end{equation}
\begin{equation}
y^{*}=\frac{\mu}{\beta}\left(\frac{N\beta}{\gamma + \mu} - 1\right),
\end{equation}
and
\begin{equation}
z^{*}=\frac{\gamma }{\beta}\left(\frac{N\beta }{\gamma + \mu } - 1\right),
\end{equation}
is a global attractor -- a stable node/focus. In the two-dimensional
invariant plane $x+y+z=N$ the basic equations of the SIR model with deaths
and births are
\begin{equation}  \label{W1}
\frac{dx}{dt}=\mu N-\beta xy-\mu x,
\end{equation}
and
\begin{equation}  \label{W2}
\frac{dy}{dt}=\beta xy-(\gamma +\mu)y,
\end{equation}
respectively. Let $x^{*}$ and $y^{*}$ the equilibrium points of the system.
In the following we rigorously show that the equilibrium $%
\left(x^{*},y^{*}\right)$ is globally asymptotically stable, i.e., all
initial conditions with $x(0) > 0$ and $y(0) > 0$ give solutions that
converge onto this equilibrium point. We will prove this result by using the
Lyapounov direct method. As a first step we scale the variables by
population size, so that $x\rightarrow x/N$, and $y\rightarrow y/N$,
respectively. Next we introduce the function $L(x,y)$ defined as \cite{Mur1}
\begin{equation}
L(x,y)=x-x^{*}\ln x+y-y^{*}\ln y, x,y\in (0,1).
\end{equation}
Then, with the use of Eqs.~(\ref{W1}) and (\ref{W2}) it immediately follows
that
\begin{equation}
\frac{dL(x,y)}{dt}=\nabla V(x,y)\cdot \left(\frac{dx}{dt},\frac{dy}{dt}%
\right)<0,x,y\in (0,1).
\end{equation}
$L(x,y)$ is therefore a Lyapunov function for the basic SIR model with vital
dynamics. The existence of a Lyapounov function $L$ ensures the global
asymptotic stability of the equilibrium point $\left(x^{*},y^{*}\right)$ %
\cite{Mur1}.

\subsection{The evolution equation of the SIR models with vital dynamics}

We derive now the basic differential equation describing the dynamics of the
SIR model with equal birth and death rates. We differentiate Eq.~(\ref{K3})
with respect to the time $t$ and obtain first the second order differential
equation
\begin{equation}
y^{\prime }=\frac{1}{\gamma }\left( z^{\prime \prime }+\mu z^{\prime
}\right) .  \label{K4}
\end{equation}

By inserting Eq.~(\ref{K3}) into Eq.~(\ref{K1}) leads to the following
differential equation,
\begin{equation}
x^{\prime }=-\frac{\beta }{\gamma }x\left( z^{\prime }+\mu z\right) +\mu
\left( N-x\right) .  \label{K5}
\end{equation}
Now, substituting Eqs.~(\ref{K3}) and (\ref{K4}) into Eq.~(\ref{K2}) yields
the differential equation:
\begin{equation}
\beta x=\frac{z^{\prime \prime }+\mu z^{\prime }}{z^{\prime }+\mu z}+\gamma
+\mu .  \label{K6}
\end{equation}
Then, by differentiating Eq.~(\ref{K6}) with respect to the time $t$ gives
the third order differential equation,
\begin{equation}
\beta x^{\prime }=\frac{z^{\prime \prime \prime }+\mu z^{\prime \prime }}{%
z^{\prime }+\mu z}-\left( \frac{z^{\prime \prime }+\mu z^{\prime }}{%
z^{\prime }+\mu z}\right) ^{2}.  \label{K7}
\end{equation}

Finally, by substituting Eqs.~(\ref{K6}) and (\ref{K7}) into Eq.~(\ref{K5})
gives the basic differential equation describing the SIR model with equal
rate of deaths and births,
\begin{equation}
\frac{z^{\prime \prime \prime }+\mu z^{\prime \prime }}{z^{\prime }+\mu z}%
-\left( \frac{z^{\prime \prime }+\mu z^{\prime }}{z^{\prime }+\mu z}\right)
^{2}=-\left( \frac{z^{\prime \prime }+\mu z^{\prime }}{z^{\prime }+\mu z}%
+\gamma +\mu \right) \left[ \frac{\beta }{\gamma }\left( z^{\prime }+\mu
z\right) +\mu \right] +\beta \mu N.  \label{K8}
\end{equation}

Eq.~(\ref{K8}) must be integrated by taking into account the initial
conditions of the SIR model with equal rates of deaths and births, given by $%
z(0)=N_3$, $z^{\prime}(0)=\gamma N_2-\mu N_3$, and $z^{\prime\prime}(0)=%
\beta \gamma N_1N_2-\gamma (\gamma +2 \mu ) N_2+\mu ^2 N_3$, respectively.

\subsection{Reduction of the evolution equation for the SIR model with vital
dynamics to an Abel type equation}

In order to simplify Eq.~(\ref{K8}), we introduce a set of transformations
defined as
\begin{equation}  \label{K91}
\psi =z^{\prime }+\mu z,
\end{equation}
\begin{equation}  \label{K92}
\psi ^{\prime }=z^{\prime \prime }+\mu z^{\prime },
\end{equation}
\begin{equation}
\psi ^{\prime \prime }=z^{\prime \prime \prime }+\mu z^{\prime \prime }.
\label{K9}
\end{equation}

By substituting Eqs.~(\ref{K91})--(\ref{K9}) into Eq.~(\ref{K8}) leads to a
second order differential equation
\begin{equation}
\left( \frac{\psi ^{\prime }}{\psi }\right) ^{2}-\frac{\psi ^{\prime \prime }%
}{\psi }=\frac{\beta }{\gamma }\psi ^{\prime }+\mu \frac{\psi ^{\prime }}{%
\psi }+\beta \left( \frac{\mu }{\gamma }+1\right) \psi +\mu \left( \mu
+\gamma -\beta N\right) ,  \label{K10}
\end{equation}
which can be rewritten as
\begin{equation}
\left( \frac{d\psi }{dt}\right) ^{2}-\psi \frac{d^{2}\psi }{dt^{2}}=\left(
\mu \psi +c\psi ^{2}\right) \frac{d\psi }{dt}+b\psi ^{2}+a\psi ^{3},
\label{K11}
\end{equation}%
where we have denoted
\begin{equation}
a=\beta \left( \frac{\mu }{\gamma }+1\right) ,
\end{equation}%
\begin{equation}
b=\mu \left( \mu +\gamma -\beta N\right) ,  \label{b}
\end{equation}%
and
\begin{equation}
c=\frac{\beta }{\gamma },
\end{equation}%
respectively. The initial conditions for the integration of Eq.~(\ref{K11})
are $\psi (0)=\gamma N_2$, and $\psi ^{\prime}(0)=\gamma \left[\beta
N_1-\left(\gamma +\mu \right)\right]N_2$.

With the help of the transformation defined as
\begin{equation}
w=\frac{dt}{d\psi }=\frac{1}{\psi ^{\prime}},  \label{w}
\end{equation}%
Eq.~(\ref{K11}) takes the form of the standard Abel type first order
differential equation of the first kind,
\begin{equation}
\frac{dw}{d\psi }=\left( a\psi ^{2}+b\psi \right) w^{3}+\left( c\psi +\mu
\right) w^{2}-\frac{1}{\psi }w,  \label{Abel}
\end{equation}
with the corresponding initial condition given by $w\left(\gamma
N_2\right)=1/\gamma \left[\beta N_1-(\gamma +\mu )\right]N_2$.

By introducing a new function $v$ defined as
\begin{equation}
v=w\psi =\frac{\psi}{\psi^{\prime}},  \label{T}
\end{equation}%
the Abel Eq. (\ref{Abel}) reduces to the form
\begin{equation}
\frac{dv}{d\psi }=\left( a+\frac{b}{\psi }\right) v^{3}+\left( c+\frac{\mu }{%
\psi }\right) v^{2},  \label{Abel2}
\end{equation}%
which is equivalent to the non-linear system of Eqs. (\ref{K1})--(\ref{K3}),
and must be integrated with the initial condition $v\left(\gamma
N_2\right)=1/\left[\beta N_1-(\gamma +\mu )\right]$.

The mathematical properties of the Abel type equation, and its applications,
have been intensively investigated in a series of papers \cite{11,12,13,
Rosu1, Rosu2}. Note that when the average death rate $\mu $ is zero, in view
of Eq.~(\ref{b}) then $b=0$, and the Abel Eq.~(\ref{Abel2}) becomes a
separate variable type differential equation of the form
\begin{equation}
\frac{dv}{d\psi }=av^{3}+cv^{2}.  \label{s}
\end{equation}

Eq.~(\ref{s}) is equivalent to the system of differential equations Eqs.~(%
\ref{A1})-(\ref{A3}) describing the SIR epidemic model without deaths. We
shall not present the simple solution of Eq.~(\ref{s}) here since we have
already presented the complete exact solution of Eqs.~(\ref{A1})-(\ref{A3})
in Section \ref{II}.

\subsection{The iterative solution of the Abel equation}

By introducing a new independent variable $\Psi $ defined as
\begin{equation}
\Psi =\ln v,
\end{equation}%
Eq.~(\ref{Abel2}) takes the form
\begin{equation}
\frac{d\Psi }{d\psi }=\left( a+\frac{b}{\psi }\right) e^{2\Psi }+\left( c+%
\frac{\mu }{\psi }\right) e^{\Psi },  \label{Abel3}
\end{equation}%
or, equivalently,
\begin{equation}
\frac{d\Psi }{d\psi }=\left( a+\frac{b}{\psi }\right) \left[ 1+2\Psi +\frac{%
\left( 2\Psi \right) ^{2}}{2!}+\frac{\left( 2\Psi \right) ^{3}}{3!}+...%
\right] +\left( c+\frac{\mu }{\psi }\right) \left[ 1+\Psi +\frac{\Psi ^{2}}{%
2!}+\frac{\Psi ^{3}}{3!}+...\right] ,  \label{49}
\end{equation}%
and must be integrated with the initial condition given by $\Psi \left(
\gamma N_{2}\right) =-\ln \left| \beta N_{1}-(\gamma +\mu )\right| $.

In the limit of small $\Psi $, in the zero order of approximation Eq.~(\ref%
{49}) becomes a first order differential equation of the form
\begin{equation}
\frac{d\Psi _{0}}{d\psi }=\left( 2a+c+\frac{2b+\mu }{\psi }\right) \Psi
_{0}+a+c+\frac{b+\mu }{\psi },
\end{equation}%
with the general solution given by
\begin{eqnarray}
\Psi _{0}(\psi ) &=&e^{(2a+c)\psi }\psi ^{2b+\mu }\left[ C_{0}+\int {%
e^{-(2a+c)\psi }\psi ^{-(2b+\mu )}\left( a+c+\frac{b+\mu }{\psi }\right)
d\psi }\right] = \nonumber\\
&&\frac{(\gamma N_{2})^{-2b-\mu }e^{-(2a+c)(\gamma N_{2}-\psi )}}{2a+c} %
\Bigg\{\psi ^{2b+\mu }\Bigg[ -(bc-a\mu )e^{\gamma N_{2}(2a+c)}E_{2b+\mu
+1}((2a+c)N_{2}\gamma )-\nonumber\\
&&(2a+c)\ln |\beta N_{1}-\gamma -\mu |+a+c\Bigg]-(\gamma N_{2})^{2b+\mu }e^{(2a+c)(\gamma N_{2}-\psi )}\times \nonumber\\
&&\left[a+c -e^{\psi
(2a+c)}(bc-a\mu )E_{2b+\mu +1}((2a+c)\psi )\right] \Bigg\},
\end{eqnarray}
where $E_{n}(x)$ is the exponential integral $E_{n}(x)=\int_{1}^{\infty }{%
e^{-xt}/t^{n}dt}$, and $C_{0}$ is an arbitrary constant of integration,
which has been determined from the initial condition.


In order to obtain the solution of the Abel equation in the next order of
approximation, we write Eq.~(\ref{49}) as
\begin{equation}
\frac{d\Psi }{d\psi }=\left( 2a+c+\frac{2b+\mu }{\psi }\right) \Psi +a+c+%
\frac{b+\mu }{\psi }+\sum_{k=2}^{\infty }{\left[ 2^{k}\left( a+\frac{b}{\psi
}\right) +\left( c+\frac{\mu }{\psi }\right) \right] \frac{\Psi ^{k}}{k!}}.
\label{52}
\end{equation}


To obtain the first order approximation $\Psi _{1}$ of the solution of the
Abel equation, we substitute in Eq.~(\ref{52}) the non-linear terms
containing $\Psi $ with $\Psi _{0}$. Therefore the first order approximation
$\Psi _{1}$ for Eq.~(\ref{49}) or Eq.~(\ref{52}) satisfies the following
linear differential equation,
\begin{equation}
\frac{d\Psi _{1}}{d\psi }=\left( 2a+c+\frac{2b+\mu }{\psi }\right) \Psi
_{1}+a+c+\frac{b+\mu }{\psi }+\sum_{k=2}^{\infty }{\left[ 2^{k}\left( a+%
\frac{b}{\psi }\right) +\left( c+\frac{\mu }{\psi }\right) \right] \frac{%
\Psi _{0}^{k}}{k!}},  \label{53}
\end{equation}
with the general solution given by
\begin{eqnarray}
\Psi _{1}(\psi )&=&e^{(2a+c)\psi }\psi ^{2b+\mu}\left\{C_0+\int{%
e^{-(2a+c)\psi }\psi ^{-(2b+\mu)}\left[a+c+\frac{b+\mu }{\psi}%
+\sum_{k=2}^{\infty }{\left[ 2^{k}\left( a+\frac{b}{\psi }\right) +\left( c+%
\frac{\mu }{\psi }\right) \right] \frac{\Psi _{0}^{k}}{k!}}\right]d\psi }%
\right\}  \notag \\
&=&\Psi _{0}(\psi )+e^{(2a+c)\psi }\psi ^{2b+\mu}\int {e^{-(2a+c)\psi }\psi
^{-(2b+\mu) }\sum_{k=2}^{\infty }{\left[ 2^{k}\left( a+\frac{b}{\psi }%
\right) +\left( c+\frac{\mu }{\psi }\right) \right] \frac{\Psi _{0}^{k}}{k!}%
d\psi }}.  \label{ww}
\end{eqnarray}

The $n$-th order of approximation of the Abel equation satisfies the
following linear differential equation,
\begin{equation}
\frac{d\Psi _{n}}{d\psi }=\left( 2a+c+\frac{2b+\mu }{\psi }\right) \Psi
_{n}+a+c+\frac{b+\mu }{\psi }+\sum_{k=2}^{\infty }{\left[ 2^{k}\left( a+%
\frac{b}{\psi }\right) +\left( c+\frac{\mu }{\psi }\right) \right] \frac{%
\Psi _{n-1}^{k}}{k!}}.
\end{equation}%
with the general iterative solution given by
\begin{eqnarray}
\Psi _{n}(\psi ) &=&e^{(2a+c)\psi }\psi ^{2b+\mu }\left\{ C_{0}+\int {%
e^{-(2a+c)\psi }\psi ^{-(2b+\mu )}\left[ a+c+\frac{b+\mu }{\psi }%
+\sum_{k=2}^{\infty }{\left[ 2^{k}\left( a+\frac{b}{\psi }\right) +\left( c+%
\frac{\mu }{\psi }\right) \right] \frac{\Psi _{n-1}^{k}}{k!}}\right] d\psi }%
\right\}  \notag \\
&=&\Psi _{0}(\psi )+e^{(2a+c)\psi }\psi ^{2b+\mu }\int {e^{-(2a+c)\psi }\psi
^{-(2b+\mu )}\sum_{k=2}^{\infty }{\left[ 2^{k}\left( a+\frac{b}{\psi }%
\right) +\left( c+\frac{\mu }{\psi }\right) \right] \frac{\Psi _{n-1}^{k}}{k!%
}d\psi }}.  \label{ww1}
\end{eqnarray}


Therefore the general solution of the Abel equation can be obtained as
\begin{equation}
\Psi (\psi )=\lim_{n\rightarrow \infty }\Psi _{n}(\psi ).
\end{equation}

Once the function $\Psi $ is known, we obtain immediately $v(\psi )=e^{\Psi
(\psi )}$, and $w(\psi )=e^{\Psi (\psi )}/\psi $, respectively. Therefore
the time evolution of the SIR model with deaths can be obtained, as a
function of the parameter $\psi $, as
\begin{equation}
t-t_{0}=\int {wd\psi }=\int {\frac{e^{\Psi (\psi )}}{\psi }d\psi }.
\label{timepar}
\end{equation}

In terms of the variable $\psi $, Eq.~(\ref{K5}) for $x$ becomes
\begin{equation}
\frac{dx}{d\psi }=-e^{\Psi (\psi )}\left( c+\frac{\mu }{\psi }\right) x+\mu N%
\frac{e^{\Psi (\psi )}}{\psi },  \label{finx}
\end{equation}%
and must be integrated with the initial condition $x\left( \gamma
N_{2}\right) =N_{1}$. Eq.~(\ref{finx}) has the general solution given by
\begin{eqnarray}\label{xpar}
x\left( \psi \right)&=& e^{-\int_1^{\psi } \frac{e^{\Psi (\xi )} (c \xi +\mu )}{\xi } \, d\xi } \Bigg[\int_1^{\psi } \frac{\mu  N e^{\Psi (\chi )+\int_1^{\chi } \frac{e^{\Psi (\xi )} (c \xi +\mu
   )}{\xi } \, d\xi }}{\chi } \, d\chi -\int_1^{\gamma  N_2} \frac{\mu  N e^{\Psi (\chi )+\int_1^{\chi } \frac{e^{\Psi (\xi )} (c \xi +\mu )}{\xi } \, d\xi }}{\chi } \,
   d\chi +\nonumber\\
 &&  N_1 e^{\int_1^{\gamma  N_2} \frac{e^{\Psi (\xi )} (c \xi +\mu )}{\xi } \, d\xi }\Bigg].
\end{eqnarray}

With the use of Eq.~(\ref{K92}), Eq.~(\ref{K4}) for $y$ takes the form
\begin{equation}
y^{\prime }=\frac{1}{\gamma }\psi ^{\prime },
\end{equation}%
with the general solution
\begin{equation}
y(\psi )=\frac{1}{\gamma }\psi +Y_{0},  \label{ypar}
\end{equation}%
where $Y_{0}$ is an arbitrary constant of integration. By estimating Eq.~(%
\ref{ypar}) at $t=0$, corresponding to $\left. \psi \right| _{t=0}=\gamma
N_{2}$, we obtain
\begin{equation}
y\left( \gamma N_{2}\right) =y(t=0)=N_{2}=N_{2}+Y_{0},
\end{equation}%
a condition that fixes the integration constant $Y_{0}$ as $Y_{0}=0$.

Finally, in the new variable $\psi $, Eq.~(\ref{K91}) for $z$ takes the form
\begin{equation}
\psi =\frac{dz}{d\psi }\frac{d\psi }{dt}+\mu z,
\end{equation}%
or, equivalently,
\begin{equation}
\frac{dz(\psi )}{d\psi }=-\mu \frac{e^{\Psi (\psi )}}{\psi }z(\psi )+e^{\Psi
(\psi )},  \label{zfin}
\end{equation}%
with the initial condition $z\left( \gamma N_{2}\right) =N_{3}$. The general
solution of Eq.~(\ref{zfin}) is provided by
\begin{eqnarray}
z(\psi )= e^{-\mu\int_1^{\psi } \frac{  e^{\Psi (\xi )}}{\xi } \, d\xi } \left[\int_1^{\psi } e^{\Psi (\chi )+\mu\int_1^{\chi } \frac{  e^{\Psi (\xi )}}{\xi } \, d\xi } \, d\chi
   -\int_1^{\gamma  N_2} e^{\Psi (\chi )+\mu\int_1^{\chi } \frac{  e^{\Psi (\xi )}}{\xi } \, d\xi } \, d\chi +N_3 e^{\mu\int_1^{\gamma  N_2} \frac{
   e^{\Psi (\xi )}}{\xi } \, d\xi }\right].  \label{zpar}
\end{eqnarray}

Eqs.~(\ref{timepar}), (\ref{xpar}), (\ref{ypar}), and (\ref{zpar}) give the
general solution of the SIR model with vital dynamics, in a parametric form,
with $\psi $ taken as a parameter.

\begin{figure}[tbp]
\centering \includegraphics[width=100mm, height=70mm]{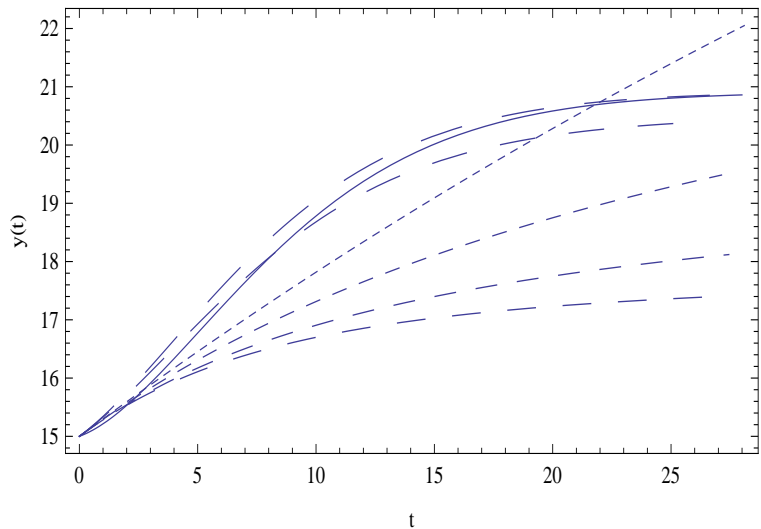}
\caption{ Comparison of $y(t)$, obtained from the numerical integration of
the equations of the SIR model with vital dynamics (solid curve), and of $%
y_n(t)$, obtained by iteratively solving the Abel Eq.~(\ref{49}), for
different orders of approximations: $n=1$ (dotted curve), $n=2$ (small
dashed curve), $n=3$ (medium dashed curve), $n=5$ (dashed curve), $n=10$
(long dashed curve), and $n=20$ (ultra-long dashed curve). The initial value
of $y(t)$ is $y(0)=15$. }
\label{fig4}
\end{figure}

In Fig.~\ref{fig4} we present the comparison of the exact numerical solution
for $y(t)$ with the different order approximations obtained by iteratively
solving the Abel Eq.~(\ref{49}). After twenty steps the iterative and the
numerical solution approximately overlap.

\section{Conclusions}

\label{IV}

In the present paper we have considered two versions of the SIR model,
describing the spread of an epidemic in a given population. For the SIR
model without births and deaths the exact analytical solution was obtained
in a parametric form. The main properties of the exact solution were
investigated numerically, and it was shown that it reproduces exactly the
numerical solution of the model equations.

For the SIR model with births and deaths we have shown that the non-linear
system of differential equations governing it can be reduced to the Abel
Eq.~(\ref{Abel2}). This Abel equation can be easily studied by means of
semi-analytical/numerical methods, thus leading to a significant
simplification in the study of the model. Once the general solution of the
Abel equation is known, the general solution of the SIR epidemic model with
deaths can be obtained in an exact parametric form.

The exact solution is important because biologists could use it to run
experiments to observe the spread of infectious diseases by introducing
natural initial conditions. Through these experiments one can learn the ways
on how to control the spread of epidemics.

\section*{Acknowledgments}

We would like to thank the two anonymous referees for comments and
suggestions that helped us to significantly improve our manuscript. We also express our special thanks to Ana Nunes for a careful reading of the manuscript, and for very helpful comments and suggestions.

\end{document}